\documentclass[reprint,superscriptaddress,onecolumn,showpacs,showkeys]{revtex4}
\usepackage{amssymb,amsmath,epsfig}
\usepackage{changes}
\usepackage{eurosym}
\usepackage{amsfonts}
\usepackage{array}
\usepackage{amsthm}
\usepackage{bm}
\usepackage{palatino}
\usepackage[colorinlistoftodos]{todonotes}
\usepackage{mathpazo}
\usepackage{amssymb}
\usepackage{eurosym}
\usepackage{amsmath}
\usepackage{epsfig}
\usepackage{graphics}
\usepackage{color}
\usepackage{graphicx}
\usepackage{hyperref}

\begin{document}

\title{\textbf{The effect of the Brane-Dicke coupling parameter on weak gravitational lensing by  wormholes and naked singularities}}

\author{Wajiha Javed} \email{wajiha.javed@ue.edu.pk;
wajihajaved84@yahoo.com}
\affiliation{Department of Mathematics, University of Education,\\
Township, Lahore-54590, Pakistan.}

\author{Rimsha Babar}
\email{rimsha.babar10@gmail.com}
\affiliation{Department of Mathematics, University of Education,\\
Township, Lahore-54590, Pakistan.}

\author{Ali \"{O}vg\"{u}n}
\email{ali.ovgun@pucv.cl}
\homepage[]{https://www.aovgun.com}
\affiliation{Instituto de F\'{\i}sica, Pontificia Universidad Cat\'olica de Valpara\'{\i}%
so, Casilla 4950, Valpara\'{\i}so, Chile}
\affiliation{Physics Department, Faculty of Arts and Sciences, Eastern Mediterranean
University, Famagusta, North Cyprus, via Mersin 10, Turkey}

%\date{\today}

\begin{abstract}
In this paper, we analyze the deflection angle of light by Brane-Dicke
wormhole in the weak field limit approximation to find the effect of the Brane-Dicke coupling parameter on the weak gravitation lensing. For this purpose, we consider new geometric techniques, i.e., Gauss-Bonnet theorem and optical geometry in order to calculate the deflection angle. Furthermore, we verify our results by considering the most familiar geodesic technique. Moreover, we establish the quantum corrected metric of Brane-Dicke wormhole by replacing the classical geodesic with Bohmian trajectories, whose matter source and anisotropic pressure are influenced by Bohmian quantum effects and calculate its
quantum corrected deflection angle. Then, we calculate the deflection angle by naked singularities and compare with the result of wormhole's.  Such a novel lensing feature might serve as a way to detect wormholes, naked singularities and also the evidence of Brane-Dicke theory.
\end{abstract}

\keywords{ Deflection of light, Gravitational lensing, Wormhole, Gauss-Bonnet theorem, Brane-Dicke gravity.}
\pacs{04.40.-b, 95.30.Sf, 98.62.Sb}

\maketitle

\section{Introduction}

Last year researchers detected the first neutron star collision \cite{TheLIGOScientific:2017qsa}. It was an epochal discovery. This is only just the beginning of gravitational wave astronomy. Since the discovery of gravitational waves, many of the modified gravity theories face difficulties to solve problems which were invented for. Furthermore, new data from the months since this discovery made life more and more difficult for the defenders of the many remaining modified-gravity theories \cite{Berti:2018cxi}. For solving this issue, physicists analyze the rotating neutron stars or black holes to find some differences between their observations and general relativity estimates - the inconsistencies predicted by some alternative gravitational theories. These systems allow astronomers to study gravity on a new scale and with new precision. In every new observation, these alternative gravitational theories are becoming increasingly difficult to solve the problems they have invented. For an alternative gravitational theory to work, it must replicate not only dark matter and dark energy, but also general relativity estimates in all standard contexts. One of the proposed the modifications of general relativity theory is the Brans-Dicke (BD) theory \cite{Brans:1961sx} which is in the category of  Galileon theories similarly to dilaton theories, chameleon theories and quintessence that attempt to get rid of dark energy and explain the expansion of the universe. There are only a few sets of dimensions in which the string theories is consistent in itself, and the most promising is four-dimensional gravity which defines our Universe \cite{Garriga:1999yh}. However, a 10-dimensional BD theory of gravity was founded and to regain the seriousness of our universe, you have to get rid of the six dimensions and take the BD coupling parameter $\omega$ which is a dimensionless constant, to infinity. The BD gravitation theory is a theoretical framework for explaining gravity (  a scalar tensor theory) where the gravitational interaction is mediated by the scalar field as well as the general relativity tensor field \cite{Hawking:1972qk}. They also predict that gravitational waves are emitted more slowly than light. Although the gravity constant G is not assumed to be constant, $\frac{1}{ G}$ is replaced by a scalar field such as $ { \phi}$, which can vary depending on location and time. The BD coupling parameter $ \omega $ can be selected to match the observations \cite{Agnese:1995kd}. The BD theory also predicts the deviation of light and the continuity of planets orbiting the Sun and the deflection angle depends on the BD  coupling constant $\omega$. This means that from the observations of the astronomical systems it is possible to put an observational lower limit to the possible $ \ omega $ value. Moreover, general relativity is derived from the Brans-Dicke theory at the boundary of omega's infinity \cite{A1}.

Wormhole; physicist Einstein's solution of the field equations in the general theory of relativity similar to a tunnel between two black holes in space-time or a tunnel between other points. Traveling in a short distances between the stars make the wormholes attractive for space travel. It was also claimed that the tunnel would allow a wormhole to travel back into the past. However, the wormholes are intrinsically unstable. Although exotic stabilization schemes have been proposed, there is no evidence for the wormholes \cite{Morris:1988cz,Morris:1988tu,Visser:1989kh,Guendelman:2009zz,Guendelman:2009pf,Guendelman:2008zp,Maldacena:2018gjk,Susskind:2017nto,Jensen:2013ora,Gao:2016bin,Konoplya:2018ala,Bronnikov:2005gm,Bronnikov:2002rn,Lobo:2005us,Clement:1983fe,Perry:1991qq,Halilsoy:2013iza,Ovgun:2018xys,Sakalli:2015taa,Richarte:2017iit,Ovgun:2015sqa}. One of the way to detect the nature of the wormhole is to use gravitational lensing \cite{Perlick:2003vg,Abe:2010ap,Safonova:2001vz,Tsukamoto:2012xs,Sharif:2015qfa,Jusufi:2018kmk}.

In this work, we study the BD wormhole solutions to investigate the gravitational lensing in weak field limits. Recently, Gibbons and Werner (GW) have proposed the method to obtain the deflection angle by black holes in the weak field limit \cite{A2}. The optical geometry is used to calculate the Gaussian curvature $\mathcal{K}$ and then using the following Gauss-Bonnet theorem, one can find the deflection angle for the asymptotically flat spacetimes:
\begin{equation}
\hat{\alpha}=-\int \int_{D_\infty} \mathcal{K}  \mathrm{d}S,
\end{equation} 
which gives exact results for bending angle in leading order terms. Then this new method by GW was applied to a variety of spacetime metric of black holes and wormholes \cite{Gibbons:2008hb,Gibbons:2008zi,Gibbons:2011rh,Bloomer:2011rd,Werner:2012rc,Gibbons:2015qja,Ishihara:2016vdc,Das:2016opi,Sakalli:2017ewb,Jusufi:2017lsl,Ono:2017pie,Jusufi:2017hed,Ishihara:2016sfv,Jusufi:2017vta,Goulart:2017iko,Jusufi:2017uhh,Arakida:2017hrm,Crisnejo:2018uyn,Jusufi:2018jof,Ovgun:2018fnk,Ovgun:2018ran,Ovgun:2018prw,Ono:2018ybw,Ovgun:2018oxk,Ovgun:2018tua,Crisnejo:2018ppm,Ono:2018jrv,Gonzalez:2018zuu}.

The main motivation of the paper is to find the effect of the BD coupling parameter on the deflection angle by wormhole. To do so, the paper is organized as follows:
In Section \textbf{II}, we investigate the deflection angle of light by BD wormhole Class II using the geodesic method and the new method by GW.
Section \textbf{III}, we study the deflection angle of light by BD wormhole Class I using the geodesic and the new method by GW, then we verify our results.
In Section \textbf{IV}, we establish the quantum
corrected metric of BD wormhole Class I as well as analyze the
quantum corrected deflection angle of light from it using the new method by GW. In Section \textbf{V}, we study the weak gravitational lensing by naked singularities. Finally, the results of this paper are
summarized in Section \textbf{VII}.

\section{Brane-Dicke Wormhole Class II}
The fields equations of the BD field is given as follows \cite{A1}:
\begin{equation}
\begin{aligned} \left( \phi ^ { ; \rho } \right) _ { ; \rho } = & \frac { 8 \pi } { 3 + 2 \omega } T _ { \mu } ^ { \mu } ,\\ R _ { \mu \nu } - \frac { 1 } { 2 } g _ { \mu \nu } R = & - \frac { 8 \pi } { \phi } T _ { \mu \nu } - \frac { \omega } { \phi ^ { 2 } } \left[ \phi _ { ; \mu } \phi _ { ; \nu } - \frac { 1 } { 2 } g _ { \mu \nu } \phi _ { ; \rho } \phi ^ { ; \rho } \right], \\ & - \frac { 1 } { \phi } \left[ \phi _ { ; \mu ; \nu } - g _ { \mu \nu } \left( \phi ^ { ; \rho } \right) _ { ; \rho } \right], \end{aligned}
\end{equation}
where $ T_{mn}$ is the energy-momentum tensor for matter with excluding the field $\phi$ field, and $\omega$ is a parameter without any dimensions. One can obtain the the metric of new BD wormhole class II using the above BD field equations as \cite{A1}

\begin{equation}
ds^2=-e^{2\Phi(\tilde{R})}dt^2+\left[1-\frac{\tilde{b}(\tilde{R})}
{\tilde{R}}\right]^{-1}d\tilde{R}^2+\tilde{R}^{2}\left[d\theta^{2}+\sin^{2}
\theta d\phi^{2}\right],\label{a23}
\end{equation}

where
\begin{eqnarray}
\Phi(\tilde{R})&=&\alpha_{\circ}=0,~~~~~~~
\tilde{R}=re^{\beta_{\circ}}(1+\tilde{B}/r)^2\left(\frac{1-\tilde{B}/r}{1+\tilde{B}/r}\right)^{1-\tilde{C}}\label{a21},\\
\beta_{\circ}&=&0,~~~~~~~\tilde{b}(\tilde{R})= \tilde{R}\left[1-\left(\frac{r^2-2\tilde{B}\tilde{C} r(\tilde{R})+\tilde{B}^2}
{r^2(\tilde{R})-\tilde{B}^2}\right)^2\right]\label{a22},
\end{eqnarray}

and
\begin{eqnarray}
{ \tilde{C} ^ { 2 } \equiv  \left( \frac { \omega + 2 } { 2 } \right) ^ { - 1 } > 0 }.
\end{eqnarray}
Note that the $\tilde{b}(\tilde{R})$ and $\Phi(\tilde{R})$ represent the shape and
redshift functions, respectively. The term of $\phi_0$ is the integration constant. The throat of the wormhole occurs at
$\tilde{R}=\tilde{R}_{0}$, i.e., $b(\tilde{R}_{0})=\tilde{R}_{0}$.
The value of $r_{0}^\pm$ can be obtained as from Eq.(\ref{a22})
\begin{equation}
r_{0}^\pm=\tilde{B}\tilde{C}\left[1\pm\left(1-\frac{1}{\tilde{C}^2}\right)^{\frac{1}{2}}\right],
\end{equation}
where $r_{0}$ represents the radius of the wormhole throat.
By using the above value of $r_{0}^\pm$, we can obtained the value of
$\tilde{R}_{0}^\pm$ from Eq.(\ref{a21}). An important feature of wormhole geometry is that
the shape function $\tilde{b}(\tilde{R})$ must satisfy the flaring-out condition
\begin{equation}
\frac{b(\tilde{R})-\tilde{R}b'(\tilde{R})}{b^2(\tilde{R})}>0,
\end{equation}
in which $b'(\tilde{R})=\frac{db}{d\tilde{R}}\leq 0$, must be satisfied at the wormhole throat. Here $\rho_{\phi} \leq 0$ so that scalar field plays the role of exotic matter at the wormhole throat, which violates the weak energy condition \cite{A1}.
It is to be noted that
$\tilde{R}\rightarrow\infty$ as $r\rightarrow\infty$
and $\frac{\tilde{b}(\tilde{R})}{\tilde{R}}\rightarrow0$ as
$\tilde{R}\rightarrow\infty$. The $\Phi(\tilde{R})$
function is zero everywhere, therefore no horizon exists \&
$\Phi(\tilde{R})\rightarrow0$ as $\tilde{R}\rightarrow\infty$.

\subsection{Deflection angle of Brane-Dicke wormhole class II using geodesic method }
In order to calculate the geodesics of BD wormhole class II,
the Lagrangian $\tilde{L}$ can be defined by the metric (\ref{a23}) through a derivative to a
parameter $\tilde{s}$ along the path as follows
\begin{eqnarray}
2\mathcal{\tilde{L}}&=&-\dot{t}^2(\tilde{s})+\frac{(r^2(\tilde{s})-\tilde{B}^2)^2}
{r^4(\tilde{s})}\left(\frac{r(\tilde{s})-\tilde{B}}{r(\tilde{s})+\tilde{B}}\right)^{-2\tilde{C}}\dot{r}^2(\tilde{s})
+r^2(\tilde{s})\left(1+\frac{\tilde{B}}{r(\tilde{s})}\right)^4\nonumber\\
&\times&\left(\frac{1-\tilde{B}/r(\tilde{s})}{1+\tilde{B}/r(\tilde{s})}\right)^{2-2\tilde{C}}\left(\dot{\theta}^2
+\sin^2\theta\dot{\phi}^2\right).~~~~~~\label{a26}
\end{eqnarray}

By following the standard procedure, setting $2\mathcal{\tilde{L}}=0$ for photons, we derive the two constants
of motion of the geodesics at the equatorial plane
$\theta=\frac{\pi}{2}$ i.e.,
\begin{eqnarray}
\mathcal{P}_\phi&=&\frac{\partial\mathcal{\tilde{L}}}{\partial\dot{\phi}}
=2r^2(\tilde{s})\left(1+\frac{\tilde{B}}{r(\tilde{s})}\right)^4
\left(\frac{1-\tilde{B}/r(\tilde{s})}{1+\tilde{B}/r(\tilde{s})}\right)^{2-2\tilde{C}}
\dot{\phi}(\tilde{s})=\ell,\label{a27}\\
\mathcal{P}_t&=&\frac{\partial\mathcal{\tilde{L}}}{\partial\dot{t}}
=-2\dot{t}(\tilde{s})=-\tilde{\varepsilon}.\label{a28}
\end{eqnarray}

Furthermore, we consider a new variable $v(\phi)$, which is associated
to the old radial coordinate as $r=\frac{1}{v(\phi)}$, which leads to the identity
\begin{equation}
\frac{\dot{r}}{\dot{\phi}}=\frac{dr}{d\phi}=-\frac{1}{v^2}\frac{dv}{d\phi}.\label{a30}
\end{equation}
For the sake of simplicity, we use the metric conditions $\tilde{\varepsilon}=1$ \& $\ell=b$ for Eqs.(\ref{a26})-(\ref{a30}) and after some
algebraic manipulations we obtain the following relation
\begin{equation}
\left(\frac{dv}{d\phi}\right)^2-\frac{(1-B^2v^2)^2}{b^2}\left(\frac{1-Bv}{1+Bv}\right)^{-2\tilde{C}}+v^2=0.
\end{equation}
After solving the above equation, we get
\begin{equation}
\left(\frac{d\phi}{dv}\right)=\frac{b\Xi}{\sqrt{(1-B^2v^2)^2-b^2v^2\Xi^2}},\label{a28}
\end{equation}
where
\begin{equation}
\Xi=\left(\frac{1-Bv}{1+Bv}\right)^{\tilde{C}}.
\end{equation}

In order to obtain the solution of differential Eq.(\ref{a28}),
we use the following relation \cite{A3}
\begin{equation}
\Delta\phi=\pi+\tilde{\alpha},
\end{equation}
where $\tilde{\alpha}$ represents the deflection angle. The deflection
angle can be obtained by following the same procedure of Ref. \cite{A4}
\begin{equation}
\tilde{\alpha}=2|\phi_{v=1/b}-\phi_{v=0}|- \pi.
\end{equation}
After solving the Eq. (\ref{a28}) and considering the leading order terms, we obtain
\begin{equation}
\tilde{\alpha}\simeq4\,{\frac {\tilde{C}B}{b}}+\mathcal{O}(B^2, \tilde{C}^2).\label{eq38}
\end{equation}
For the value of $\tilde{C}$ the deflection angle can be written as follows
\begin{equation}
\tilde{\alpha}\simeq8\,{\frac {B}{b\sqrt {2\,\omega+4}}}+\mathcal{O}(B^2, \tilde{C}^2).\label{A14}
\end{equation}

\subsection{Deflection angle of Brane-Dicke wormhole class II using the new method}
Now we calculate the deflection angle of Brane-Dicke wormhole class II using the Gauss-Bonnet theorem \cite{A2}. The optical metric of Eq. (\ref{a23}) corresponding to the given shape
function $\tilde{b}(\tilde{R})$ is given as follows
\begin{equation}
dt^2=\frac{(r^2-\tilde{B}^2)^2}{r^4}\left(\frac{r-\tilde{B}}{r+\tilde{B}}\right)^{-2\tilde{C}}
dr^2+r^2\left(1+\frac{\tilde{B}}{r}\right)^4\left(\frac{1-\tilde{B}/r}{1+\tilde{B}/r}\right)^{2-2\tilde{C}}d\phi^2.\label{a24}
\end{equation}

One can rewrite the above metric as follows
\begin{equation}
dt^2=\tilde{h}_{\mu\nu}d\lambda^{\mu}d\lambda^{\nu}=dv^2+\xi^2(v)d\phi^2.
\end{equation}
It is to be noted that $(\mu,\nu=r,\phi)$ and $\tilde{h}=\det\tilde{h}_{\mu\nu}$.
The Gaussian optical curvature can be given: \cite{Jusufi:2018kmk}
\begin{equation}
\mathcal{\tilde{K}}=\frac{R_{icciScalar}}{2}=-\frac{1}{\xi(v)}\left[\frac{dr}{dv}\frac{d}{dr}\left(\frac{dr}{dv}\right)
\frac{d\xi}{dr}+\left(\frac{dr}{dv}\right)^2\frac{d^2\xi}{dr^2}\right].\label{a4}
\end{equation}
Using Eq. (\ref{a4}), we obtain the Gaussian optical curvature from Eq. (\ref{a24}) as follows
\begin{equation}
%\mathcal{\tilde{K}}\simeq\left(\frac{r-\tilde{B}}{r+\tilde{B}}\right)^{2\tilde{C}}
%\left(\frac{8\tilde{B}^2}{r^3}-\frac{2\tilde{B}\tilde{C}}{r^4}\right)
\mathcal{\tilde{K}}\simeq -2\,{\frac {\tilde{C}B}{{r}^{3}}}+\mathcal{O}(B^2, \tilde{C}^2).\label{a25}
\end{equation}

In order to calculate the deflection angle, we consider the straight line approximation as
$r(\phi)=\frac{b}{\sin\phi}$. Thus the deflection angle can be calculated from given equation \cite{Jusufi:2018kmk}
\begin{equation}
\tilde{\alpha}=-\int_0^{\pi}\int_{\frac{b}{\sin\phi}}^{\infty}
\mathcal{\tilde{K}}d\tilde{\sigma}.\label{a99}
\end{equation}
After substituting Eq. (\ref{a25}) into Eq. (\ref{a99}), we obtain
\begin{equation}
\tilde{\alpha}=-\int_0^{\pi}\int_{\frac{b}{\sin\phi}}^{\infty}
 \left( -2\,{\frac {\tilde{C}B}{{r}^{3}}} \right)  \left( 2\,\tilde{C}B+r \right) 
dr d\phi.
\end{equation}
The above equation implies
\begin{equation}
%\tilde{\alpha}\simeq\frac{16\tilde{B}^2}{b}+\frac{\tilde{B}\tilde{C}\pi}{2b^2}.
\tilde{\alpha}\simeq4\,{\frac {\tilde{C}B}{b}}+\mathcal{O}(B^2, \tilde{C}^2).\label{A13}
\end{equation}

Hence, we find the same deflection angle as in Eq. (\ref{eq38}) so, we get the same deflection angle as in Eq. (\ref{A14}) for the value of $\tilde{C}$.

\section{Brane-Dicke Wormhole Class I}
The metric of new BD wormhole class II is defined as \cite{A1}
\begin{equation}
ds^2=-e^{2\Phi(\tilde{R})}dt^2+\left[1-\frac{\tilde{b}(\tilde{R})}
{\tilde{R}}\right]^{-1}d\tilde{R}^2+\tilde{R}^{2}\left[d\theta^{2}+\sin^{2}
\theta d\phi^{2}\right],\label{a3}
\end{equation}

where
\begin{eqnarray}
\Phi(\tilde{R})&=&0,~~~~~~~
\tilde{R}=r\left(1+\frac{\tilde{B}^2}{r^2}\right)\exp\left[1-\frac{2}{\pi}
\arctan\left(\frac{r}{\tilde{B}}\right)\right]\tilde{\beta}_{0},\label{a2}\\
\tilde{b}(\tilde{R})&=& \tilde{R}\left[1-\left(1-\frac{2\tilde{B}(\tilde{C} r(\tilde{R})+\tilde{B})}
{r^2(\tilde{R})+\tilde{B}^2}\right)^2\right]\label{a1},
\end{eqnarray}

and

\begin{eqnarray}   { \tilde{C} ^ { 2 } \equiv - \left( \frac { \omega + 2 } { 2 } \right) ^ { - 1 } > 0 }. \end{eqnarray}

\subsection{Deflection angle of Brane-Dicke wormhole class I using geodesic method}
The Lagrangian of metric (\ref{a3}) can be defined as follows
\begin{eqnarray}
2\mathcal{\tilde{L}}&=&\left[\frac{\tilde{\beta}_{0}(r^2(\tilde{s})
+\tilde{B}^2)\left(-2\tilde{B}r(\tilde{s})
+\pi r^2(\tilde{s})-\tilde{B}^2\pi\right)e^{\left(
\frac{2}{\pi}\arctan\left[\frac{\tilde{B}}{r(\tilde{s})}\right]
\right)}}{\pi r^2(\tilde{s})\left((r^2(\tilde{s})
+\tilde{B}^2)+2\tilde{B}(\tilde{C}r(\tilde{s})+\tilde{B})\right)}
\right]^2\dot{r}^2(\tilde{s})\nonumber\\
&+&\left[\frac{\tilde{\beta}_{0}\left(r^2(\tilde{s})+\tilde{B}^2\right)}{r(\tilde{s})}e^{\left(
\frac{2}{\pi}\arctan\left[\frac{\tilde{B}}{r(\tilde{s})}\right]
\right)}~\right]^2\left(\dot{\theta}^2(\tilde{s})
+\sin^2\theta\dot{\phi}^2(\tilde{s})\right)-\dot{t}^2(\tilde{s}).~~~~~~\label{a11}
\end{eqnarray}
Setting $2\mathcal{\tilde{L}}=0$ for photons and considering the deflection of planar photons at the equatorial plane
$\theta=\frac{\pi}{2}$, two constants
of motion of the geodesics are derived at the equatorial plane
$\theta=\frac{\pi}{2}$ i.e., \cite{A3}
\begin{eqnarray}
\mathcal{P}_\phi&=&\frac{\partial\mathcal{\tilde{L}}}{\partial\dot{\phi}}=
2\left[\frac{\tilde{\beta}_{0}\left(r^2(\tilde{s})+\tilde{B}^2\right)}{r(\tilde{s})}e^{\left(
\frac{2}{\pi}\arctan\left[\frac{\tilde{B}}{r(\tilde{s})}\right]
\right)}~\right]^2\dot{\phi}^2(\tilde{s})=\ell,\label{a12}\\
\mathcal{P}_t&=&\frac{\partial\mathcal{\tilde{L}}}{\partial\dot{t}}
=-2\dot{t}(\tilde{s})=-\tilde{\varepsilon}.\label{a13}
\end{eqnarray}
Furthermore, we consider a new variable $v(\phi)$, which is associated
to the old radial coordinate as $r=\frac{1}{v(\phi)}$, which leads to the identity
\begin{equation}
\frac{\dot{r}}{\dot{\phi}}=\frac{dr}{d\phi}=-\frac{1}{v^2}\frac{dv}{d\phi}.\label{a14}
\end{equation}
For the sake of simplicity, we use the metric conditions $\tilde{\varepsilon}=1$ \& $\ell=b$ for Eqs.(\ref{a11})-(\ref{a14}) and after some
algebraic manipulations we obtain the following relation
\begin{eqnarray}
&0&=\left[\frac{\beta_0e^{\left(
\frac{2}{\pi}\arctan\left[\frac{\tilde{B}}{r}\right]
\right)^2}\left(1+B^2v^2\right)\left(-2Bv+\pi-B^2v^2\pi\right)}
{\pi\left(1+B^2v^2+2B(\tilde{C}v+Bv^2)\right)v^2}\right]^2\left(\frac{dv}{d\phi}\right)^2\nonumber\\
&-&\frac{1}{b^2}\left[\frac{\beta_0e^{\left(
\frac{2}{\pi}\arctan\left[\frac{\tilde{B}}{r}\right]
\right)^2}\left(1+B^2v^2\right)}{v}\right]^4+\left[\frac{\beta_0e^{\left(
\frac{2}{\pi}\arctan\left[\frac{\tilde{B}}{r}\right]
\right)^2}
\left(1+B^2v^2\right)}{v}\right]^2.\nonumber\\
\end{eqnarray}
The above equation implies
\begin{eqnarray}
\frac{d\phi}{dv}=\pm\frac{b\left(1+B^2v^2\right)\left(2Bv+\pi(-1
+B^2v^2)\right)}{\pi\left(1+2B\tilde{C}v+3B^2v^2\right)\sqrt{-b^2v^2+e^{\left(
\frac{2}{\pi}\arctan\left[\frac{\tilde{B}}{r}\right]
\right)^2}(1+B^2v^2)\beta_0^2}}.\label{b15}
\end{eqnarray}

By using the relation from Eq.(\ref{b15}), we can derive the following deflection angle
\begin{equation}
\tilde{\alpha}\simeq \frac{4B-8B\tilde{C}\pi}{b}+\mathcal{O}(B^2, \tilde{C}^3). \label{eq17}
\end{equation}
For the value of $\tilde{C}$ the equation can be written as
\begin{equation}
\tilde{\alpha}\simeq 4\,{\frac {B}{b}}-8\,{\frac {\pi\,B}{b}}+2\,{\frac {\omega\,\pi\,B}{b
}}-\,{\frac {{3\omega}^{2}\pi\,B}{4b}}+\mathcal{O}(B^2, \tilde{C}^2).\label{A11}
\end{equation}

\subsection{Deflection angle of Brane-Dicke wormhole class I using the new method}

In order to find the wormhole optical metric, we consider $ds^2=0$.
In the equatorial plane we suppose $\theta=\frac{\pi}{2}$ and
$d\theta=0$ and using Eq.(\ref{a2})-(\ref{a1}) in Eq.(\ref{a3}), we get
\begin{eqnarray}
dt^2&=&\left[\frac{\tilde{\beta}_{0}(r^2+\tilde{B}^2)\left(-2\tilde{B}r
+\pi r^2-\tilde{B}^2\pi\right)e^{\left(
\frac{2}{\pi}\arctan\left[\frac{\tilde{B}}{r}\right]
\right)}}{\pi r^2\left((r^2+\tilde{B}^2)+2\tilde{B}(\tilde{C}r+\tilde{B})\right)}\right]^2dr^2\nonumber\\
&+&\left[\frac{\tilde{\beta}_{0}\left(r^2+\tilde{B}^2\right)}{r}e^{\left(
\frac{2}{\pi}\arctan\left[\frac{\tilde{B}}{r}\right]
\right)}~\right]^2d\phi^2.
\end{eqnarray}

We calculate the Gaussian optical curvature of wormhole,
as follows:
\begin{eqnarray}
\mathcal{\tilde{K}}&=\frac{R_{icciScalar}}{2}=&\frac{-e^{\left(
\frac{2}{\pi}\arctan\left[\frac{\tilde{B}}{r}\right]
\right)^2}~}{\pi^3\beta_{0}^2}
\left[\frac{2B\pi-4B\tilde{C}\pi^2}{r^3}\right]+\mathcal{O}(B^2, \tilde{C}^2).\label{a8}
\end{eqnarray}

In order to calculate the deflection angle, we consider the straight line approximation as
$r(\phi)=\frac{b}{\sin\phi}$. Thus the deflection angle can be calculated from given equation \cite{Jusufi:2018kmk}
\begin{equation}
\tilde{\alpha}=-\int_0^{\pi}\int_{\frac{b}{\sin\phi}}^{\infty}
\mathcal{\tilde{K}}d\tilde{\sigma}.\label{a9}
\end{equation}
By putting Eq.(\ref{a8}) into (\ref{a9}), we get
\begin{eqnarray}
\tilde{\alpha}&=&-\int_0^{\pi}\int_{\frac{b}{\sin\phi}}^{\infty}\frac{-e^{\left(
\frac{2}{\pi}\arctan\left[\frac{\tilde{B}}{r}\right]
\right)^2}~}{\pi^3\beta_{0}^2}
\left[\frac{2B\pi-4B\tilde{C}\pi^2}{r^3} \right]\sqrt{\det\tilde{h}_{\mu\nu}}dr d\phi.
\end{eqnarray}
The above equation implies
\begin{equation}
\tilde{\alpha}\simeq \frac{4B-8B\tilde{C}\pi}{b}+\mathcal{O}(B^2, \tilde{C}^2).\label{A12}
\end{equation}
which is the same as in Eq. (\ref{eq17}). Thus, for the value of $\tilde{C}$,  we can also find the same result as in Eq. (\ref{A11}).

\section{Brane-Dicke Wormhole with Quantum Corrections to Class I}
In this section we analyze the BD wormhole metric (\ref{a3}) with the effects of quantum corrections.
For this purpose we focus on the quantum correction effects. The Einstein's equations
in the energy momentum tensor remain the same under the influence of quantum corrections.
\begin{equation}
\tilde{G}_{ab}=\hat{R}_{ab}-\frac{1}{2}\tilde{g}_{ab}\hat{R}
=8\pi\mathrm{\tilde{T}}^{eff}_{ab},\label{a16}
\end{equation}
where $\mathrm{\tilde{T}}^{eff}_{ab}$ and $\tilde{G}_{ab}$ denote the effective
energy-momentum tensor and Einstein tensor, respectively.
The $\mathrm{\tilde{T}}^{eff}_{ab}$ can be defined as follows
\begin{equation}
\mathrm{\tilde{T}}^{eff}_{ab}=\mathrm{\tilde{T}}_{ab}
+\mathrm{\tilde{T}}^{corr.}_{ab}.
\end{equation}
here $\mathrm{\tilde{T}}_{ab}$ represents the energy momentum tensor which can be expressed as
\begin{equation}
\mathrm{\tilde{T}}^{a}_{b}=\left(-\tilde{\rho}, \tilde{P}_r, \tilde{P}_\theta, \tilde{P}_\phi\right),
\end{equation}
where $\tilde{\rho}$ stands for energy density and $\tilde{P}_r, \tilde{P}_\theta,
\tilde{P}_\phi$ shows the non-zero components of diagonal terms.
The term $\mathrm{\tilde{T}}^{corr.}_{ab}$ can be given as
\begin{equation}
\mathrm{\tilde{T}}^{corr.}_{ab}=\left(-\tilde{\rho}^{(corr.)},
\tilde{P}_r^{(corr.)}, \tilde{P}_\theta^{(corr.)}, \tilde{P}_\phi^{(corr.)}\right).
\end{equation}
According to the definition of $4$-momentum we have
\begin{equation}
\tilde{p}_i=\tilde{\hbar}\partial_i\tilde{S},
\end{equation}
by using the wave function solution the geodesic equation can be
modified due to the relativistic quantum potential which can be defined by \cite{Jusufi:2018kmk,p1}
\begin{equation}
\tilde{V}_Q=\tilde{\hbar}^2\frac{\square\mathcal{R}}{\mathcal{R}}.
\end{equation}
The non-zero components of stress-energy tensor can be given as \cite{p1}
\begin{equation}
\tilde{\rho}^{(corr.)}=\tilde{P}_r^{(corr.)}= \tilde{P}_\theta^{(corr.)}=
\tilde{P}_\phi^{(corr.)}=\frac{\tilde{\hbar}\tilde{\eta}}{8\pi \tilde{R}^4},
\end{equation}
here $\tilde{\eta}$ stands for dimensionless constant.
The non-zero components from Einstein tensor $\tilde{G}_{ab}$
of the metric (\ref{a3}) are given as 
\begin{eqnarray}
\tilde{G}^t_t&=&-\frac{b'(\tilde{R})}{\tilde{R}^2},~~~~~~~~~~~~~~~~~~
\tilde{G}^r_r=-\frac{b(\tilde{R})}{\tilde{R}^3}+2\left(1-
\frac{b(\tilde{R})}{\tilde{R}}\right)\frac{\Phi'}{\tilde{R}},\nonumber\\
\tilde{G}^\theta_\theta&=&\left(1-\frac{b(\tilde{R})}{\tilde{R}}
\right)\left[\Phi^{''}+(\Phi')^2-\frac{b'R-b}{2R(R-b)}\Phi'
-\frac{b'R-b}{2R^2(R-b)}\Phi'+\frac{\Phi'}{R}\right],\nonumber\\
\tilde{G}^\phi_\phi&=&\tilde{G}^\theta_\theta.
\end{eqnarray}
By utilizing the metric (\ref{a3}) and Eq.(\ref{a16}), we get
the following set of Einstein field equations
\begin{eqnarray}
\tilde{\rho}(\tilde{R})&=&\frac{1}{8\pi\tilde{R}^2}\left[b'(\tilde{R})
-\frac{\tilde{\hbar}\tilde{\eta}}{\tilde{R}^2}\right],\nonumber\\
\tilde{P}_r(\tilde{R})&=&\frac{1}{8\pi}\left[2\left(1-
\frac{b(\tilde{R})}{\tilde{R}}\right)\frac{\Phi'}{\tilde{R}}
-\frac{b(\tilde{R})}{\tilde{R}^3}-\frac{\tilde{\hbar}
\tilde{\eta}}{\tilde{R}^4}\right],\nonumber\\
\tilde{P}_\theta(\tilde{R})&=&\tilde{P}_\phi(\tilde{R})
=\frac{1}{8\pi}\left(1-\frac{b(\tilde{R})}{\tilde{R}}
\right)\left[\Phi^{''}+(\Phi')^2-\frac{b'R-b}{2R(R-b)}\Phi'\right.\nonumber\\
&-&\left.\frac{b'R-b}{2R^2(R-b)}\Phi'+\frac{\Phi'}{R}
-\frac{\tilde{\hbar}\tilde{\eta}}{\tilde{R}^4}\right],
\end{eqnarray}
The geometry of wormhole under the influence of quantum effects
can be derived by shifting the shape function as follows
\begin{equation}
b\rightarrow b_{eff}=b-\frac{\tilde{\hbar}\tilde{\eta}}{\tilde{R}}.
\end{equation}
Also
\begin{equation}
\Phi(\tilde{R})\rightarrow \tilde{\Upsilon}(\tilde{R})=\frac{1}{2}
\ln\left(1+\frac{\tilde{\hbar}\tilde{\eta}}{\tilde{R}^2}\right).
\end{equation}
The Einstein's field equations in terms of effective shape
function can be expressed as follows
\begin{eqnarray}
\tilde{\rho}(\tilde{R})&=&-\frac{b'_{eff}(\tilde{R})}{8\pi\tilde{R}^2}
,~~~~~~\tilde{P}_r(\tilde{R})=\frac{1}{8\pi}\left[2\left(1-
\frac{b_{eff}(\tilde{R})}{\tilde{R}}\right)\frac{\tilde{\Upsilon}'}{\tilde{R}}
-\frac{b_{eff}(\tilde{R})}{\tilde{R}^3}\right],\nonumber\\
\tilde{P}_\theta(\tilde{R})&=&\tilde{P}_\phi(\tilde{R})=\frac{1}{8\pi}
\left(1-\frac{b_{eff}(\tilde{R})}{\tilde{R}}
\right)\left[\tilde{\Upsilon}^{''}+(\tilde{\Upsilon}')^2-\frac{b_{eff}'
R-b_{eff}}{2R(R-b_{eff})}\tilde{\Upsilon}'\right.\nonumber\\
&-&\left.\frac{b_{eff}'R-b_{eff}}{2R^2(R-b_{eff})}\tilde{\Upsilon}'
+\frac{\tilde{\Upsilon}'}{R}\right],
\end{eqnarray}
where
\begin{eqnarray}
\tilde{\Upsilon}'&=&-\frac{\tilde{\hbar}\tilde{\eta}}{\tilde{R}^3}
\left(1+\frac{\tilde{\hbar}\tilde{\eta}}{\tilde{R}^2}\right)^{-1},\nonumber\\
\tilde{\Upsilon}^{''}&=&\frac{\tilde{\hbar}\tilde{\eta}}{\tilde{R}^4}
\left(1+\frac{\tilde{\hbar}\tilde{\eta}}{\tilde{R}^2}\right)^{-1}\left[3
-\frac{2\tilde{\hbar}\tilde{\eta}}{\tilde{R}^2}\left(1+\frac{\tilde{\hbar}
\tilde{\eta}}{\tilde{R}^2}\right)^{-1}\right].
\end{eqnarray}
The spacetime metric (\ref{a3}) in-term of quantum corrections can be rewritten as follows
\begin{equation}
ds^2=-\left(1+\frac{\tilde{\hbar}
\tilde{\eta}}{\tilde{R}^2}\right)dt^2+\left[1-\frac{\tilde{b}(\tilde{R})}
{\tilde{R}}+\frac{\tilde{\hbar}
\tilde{\eta}}{\tilde{R}^2}\right]^{-1}d\tilde{R}^2+\tilde{R}^{2}\left[d\theta^{2}+\sin^{2}
\theta d\phi^{2}\right].\label{a17}
\end{equation}
\subsection{Gravitational Lensing using the Gauss-Bonnet theorem}
In this subsection we study the deflection of light for quantum
corrected metric of BD Wormhole given in Eq.(\ref{a17}).
By following the same procedure as section \textbf{2}, the Gaussian optical
curvature with quantum corrections can be calculated as follows
\begin{equation}
\mathcal{\tilde{K}}=
\left(\frac{-e^{\left(
\frac{2}{\pi}\arctan\left[\frac{\tilde{B}}{r}\right]
\right)^2}~}{\pi^3\beta_{0}^2}\right)\left[\frac{2B\pi-4B\tilde{C}h\eta\pi^2}{r^3}\right]+\mathcal{O}(B^2, \tilde{C}^2).~~\label{b8}
\end{equation}

The asymptotic form of the geodesic curvature is defined as
\begin{equation}
\lim_{\mathcal{R}}\rightarrow\frac{1}{\mathcal{R}},
\end{equation}
which yields the deflection angle as follows
\begin{equation}
\tilde{\alpha}\simeq \frac{4B-8B\tilde{C}h\eta\pi}{b}+\mathcal{O}(B^2, \tilde{C}^2).
\end{equation}
Hence we find as follows:
\begin{equation}
\tilde{\alpha}\simeq 4\,{\frac {B}{b}}-8\,{\frac {\pi\,Bh\eta}{b}}+2\,{\frac {\omega\,\pi\,Bh\eta}{b
}}-\,{\frac {{3\omega}^{2}\pi\,Bh\eta}{4b}}+\mathcal{O}(B^2, \tilde{C}^2). \end{equation}

\section{Deflection of Light by Naked Singularities in Brane-Dicke theory}
The metric of BD wormhole found firstly by Agnese and Camera is defined in the form \cite{Agnese:1995kd}
\begin{equation}
ds^2=-e^{2\Phi(\tilde{R})}dt^2+\left[1-\frac{\tilde{b}(\tilde{R})}
{\tilde{R}}\right]^{-1}d\tilde{R}^2+\tilde{R}^{2}\left[d\theta^{2}+\sin^{2}
\theta d\phi^{2}\right],\label{a233}
\end{equation}
where
\begin{eqnarray}
2\Phi(\tilde{R})=\sqrt { \frac { 2 } { 1 + \gamma } } \ln \left[ 1 - \frac { 2 \eta } { r ( \tilde{R} ) } \right] ,
\tilde{R}=r \left[ 1 - \frac { 2 \eta } { r } \right] ^ { [ 1 - \gamma \sqrt { 2 / ( 1 + \gamma ) ] } / 2 },\\
\frac{\tilde{b}(\tilde{R})}{\tilde{R}}=  1 - \frac { \{ 1 - [ \eta / r ( \tilde{R} ) ] [ 1 + \gamma \sqrt { 2 / ( 1 + \gamma ) } \} \} ^ { 2 } } { 1 - 2 \eta / r ( \tilde{R} ) }.
\end{eqnarray}

The value of $r_{0}^\pm$ can be obtained as follows:
\begin{equation}
r_{0}^\pm=r \left[ 1 + \gamma \sqrt { \frac { 2 } { 1 + \gamma } } \right],
\end{equation}
here $r_{0}$ is the radius of the wormhole throat.

In the case $\gamma<1$, there is a singularity at $\tilde{R} = 0$ and this point stands for the naked singularity. On the other hand at the limit of $\gamma \rightarrow 1$,
the BD reduces to Einstein theory where the event horizon is at $\tilde{R} = 2M$, that shows the black hole solution  shown by Hawking \cite{Hawking:1972qk}. Moreover, there is a static wormhole solution in the BD theory for the case $\gamma >1$.

Using the same procedure, first we obtain the optical metric of te naked singularity case and then calculate the Gaussian curvature in leading order as follows:

\begin{equation}
\mathcal{\tilde{K}}\simeq {\frac {\sqrt {2}\eta}{{r}^{3} \left( 2\,\sqrt {2}\gamma+1 \right) ^{2}}}+\mathcal{O}(\eta^2).
\end{equation}
Then using the GBT within the optical naked singularity spacetime (for the case $\gamma <1$), we obtain the deflection angle in weak field limits as follows:
\begin{equation}
\tilde{\alpha}=-2\,{\frac {\sqrt {2}\eta}{b}}-1/2\,{\frac {{\eta}^{2}\pi}{{b}^{2}}}+2
\,{\frac {\eta\, \left( \sqrt {2}\pi\,\eta+8\,b \right) \gamma}{{b}^{2}}}.
\end{equation}

Note that for the case of $\gamma >1$, one can find the deflection angle of the BD wormhole in weak field limits.

\section{Conclusion}
In this work, we have analyzed the deflection angle for the BD wormhole.
For this purpose, by using the
new geometric techniques founded by GW, i.e., Gauss-Bonnet theorem and optical geometry, we
have computed the deflection angle for BD wormhole.
It is also important to note that the deflection of a
light ray is calculated by outside of the lensing area
which shows that the gravitational lensing effect
is a global and even topological effect, i.e., there are more than one
light ray converging between the source and observer.
Furthermore, we have also calculated the same result of
deflection angle by using the standard geodesic technique.
Moreover, we have established the quantum corrected metric of BD wormhole by replacing the classical
geodesic with Bohmian trajectories, whose matter source and anisotropic pressure
are influenced by Bohmian quantum effects. We have also studied the
quantum corrected deflection angle for BD wormhole. Furthermore, we have investigated the weak gravitational lensing by naked singularities in BD theory and showed the difference between the deflection angle of naked singularities and wormholes.  In the BD theory, the role
of exotic matter is important for the the mass distribution in the Universe if $\gamma>1$ (or $\omega<-2)$.

 Clearly, the agreement
has been shown to arise from the GBT and the geodesics method for calculating the deflection angle in weak field limits. This method
as a quantitative tool can be used in any asymptotically flat spacetimes.

The main message of this article is that, these results
provide an excellent tool to direct detect the nature of the wormholes and naked singularities and also to find the evidence of the Brane-Dicke theory.

\acknowledgments
Authors would like to thank the referees and the editor for their valuable comments and suggestions. This work was supported by the Chilean FONDECYT Grant No. 3170035 (A\"{O}).

%\listofchanges
\end{document}